# Dynamic Network: Graphical Deformation of Penetrated Objects [1]

Ehsan Arbabi

*École Polytechnique Fédérale de Lausanne, Switzerland*
*ehsan.arbabi@a3.epfl.ch*

Abstract: In a computer-based virtual environment, objects may collide with each other. Therefore, different algorithms are needed to detect the collision and perform a correct action in order to avoid penetration. Based on the application and objects physical characteristics, a correct action can include separating or deforming the penetrated objects. In this article, by using the concepts of dynamic networks and simple physics, a method for deforming two penetrated 3D objects is proposed. In this method, we consider each primitive of the objects as an element interacting with the other elements in a dynamic network. These kinds of interactions make the elements impose force on each other and change their position, until a force-balance happens. The proposed method is implemented and tested on 3D sample objects, and the resulted deformation proved to be visually satisfying.

## 1 Introduction

One of the main concerns in computer graphics is to deform virtual objects after collision. In a dynamic computer environment, objects may collide with each other. Some algorithms are designed to detect the collision [1]. After collision detection, based on the application and type of the objects, another algorithm should be used to either separate the objects and/or deform them [2]. Depending on the applications, we may need either visually correct deformation or physically correct deformation. Although physically correct deformation also results in a visually correct deformation, but it usually needs more complex algorithms. In fact, in some applications, such as some cartoons or games, we do not need to deform the objects based on exactly correct physics laws. The only concern is to make the user satisfied by seeing such kind of deformations. But in some other applications, such mechanical simulations or biomedical application [3], the user is interested in mechanically correct deformation based on physics laws.

There are several methods for responding to the collision in computer graphics. In some cases, the collision response algorithm is considering the dynamic behavior of the objects (e.g. velocity,

---

acceleration). In some other cases, the positions of the objects are fixed and we only need to deform them in their current fixed position [2, 4].

In computer graphics, the virtual objects are mostly presented by set of vertices (in 3D space) and the covering polygons. In fact, the polygons clarify the connection among the vertices on the surface. Figure 1 shows a sphere presented by set of vertices and triangles.

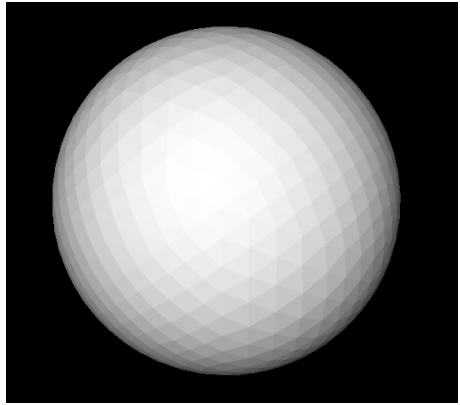

Figure 1: A sphere presented by set of vertices and triangles.

As it was explained, to avoid virtual objects from penetration, we need to 1- detect the collision and 2- deform the objects in the physically correct way. The first step provides us the penetrating vertices of the objects. In the second step, we should displace the penetrating vertices in the way that the below conditions meet (see Figure 2):

1- The objects do not penetrate any more,

2- The forces imposed on the vertices become balanced.

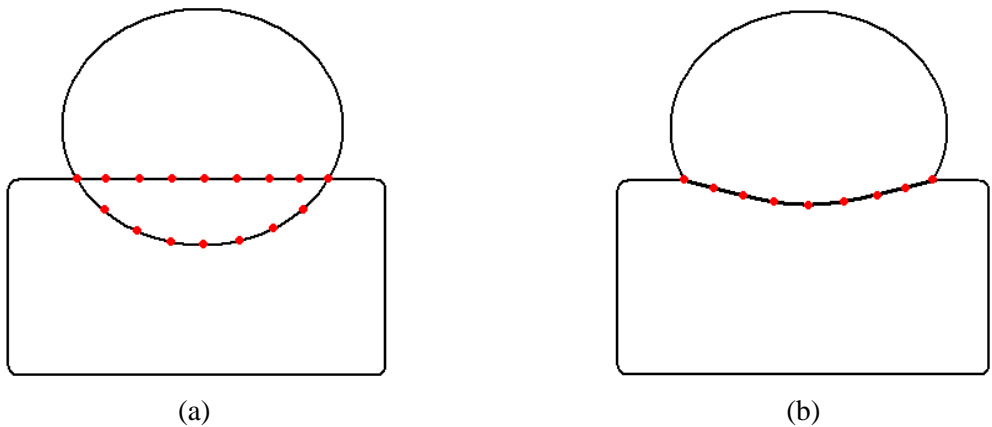

(a)                                                                (b)

Figure 2: An example for deforming two penetrating objects: (a) before deformation (unsolved penetration), (b) after deformation (no penetration). Red dots are the (a) penetrating or (b) contacting vertices.

The goal of this study is to use the concepts of the Dynamic Network, to deform two penetrating objects in their fixed position, based on simple physics laws. Although we are considering the objects fixed in their position, but the force-interaction between the objects primitives, have a dynamic behavior. Therefore, we can consider each primitive of the objects as an element interacting with the other elements in a dynamic network. These kinds of interactions make the elements impose force on each other and displace themselves, until a force-balance happens.

In this article, we first define a dynamic network based on physical interaction between two penetrated virtual objects. In the next section, the details related to implementation and testing the proposed dynamic network is presented. Finally, we conclude based on the obtained results.

## 2 Defining the dynamic network

### 2.1 Overview

After any virtual penetration, some vertices of an object go inside the other objects. Therefore, these penetrated vertices should be moved in a correct direction to solve the penetration. On the other hand, any movement of a vertex makes its neighboring vertices impose some forces on it. The forces applied on a vertex by its neighbors can be formulated as below in a simple isotropic case (by ignoring the other complexities):

$$\begin{bmatrix} \epsilon_x \\ \epsilon_y \\ \epsilon_z \end{bmatrix} = (\frac{1}{E}) \begin{bmatrix} 1 & -\nu & -\nu \\ -\nu & 1 & -\nu \\ -\nu & -\nu & 1 \end{bmatrix} \begin{bmatrix} \sigma_x \\ \sigma_y \\ \sigma_z \end{bmatrix}$$

where $\nu, E, \epsilon_i$ and $\sigma_i$ are Poisson's ratio, Young's modulus, the strain in i direction and the stress in i direction. Of course, in a much simpler case, we can consider that the objects are made up of several masses (including the vertices) connected to each other by springs.

Thus, after any displacement, we will have some forces trying to move back the vertices to their original positions. The displacement must be done in a way that the forces imposed on the vertices of an object become negated by the forces imposed by the contacting vertices of the other object(s) (as there are at least two objects penetrating each other). Any displacement of a vertex influences the forces imposed on it and its neighboring vertices. Therefore, for deforming the objects in a correct way, all the vertices and the interaction among them must be evaluated together in a dynamic network.

### 2.2 External function

As we have (at least) two objects which are penetrating each other, we also have the penetrating vertices and non-penetrating vertices (in all the objects). For each penetrating vertex in object 1, there is at least one corresponding penetrating vertex in object 2 which has the shortest distance to it (usually shortest, but not always). Two corresponding vertices and their penetration depth are shown in Figure 3.

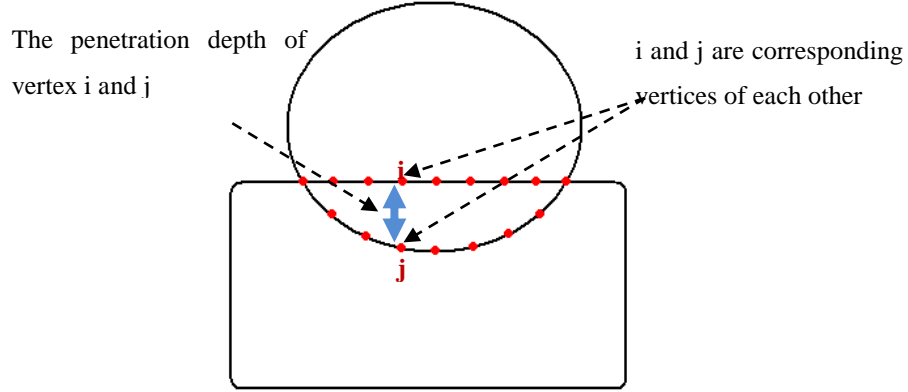

Figure 3: Corresponding vertices and their penetration depth.

The information about the corresponding vertices is provided by an external function of Collision Detection. If A is the set of vertices in 3D space:

$$A = \{a_1, a_2, \ldots, a_N\}; \text{ N: total number of vertices}$$

There is a function

$$Cv: A \rightarrow A^m; \ 0 \leq m < N.$$

The external function of Collision detection, i.e. CollisionDetection(), takes A as input and provides Cv(A) as outputs:

$$Cv(A) = \text{CollisionDetection}(A)$$
$$Cv(a_k) = \{a_i, a_j, \ldots\}; \ (a_i, a_j \text{ are the corresponding vertices of } a_k).$$

Therefore by using Cv(), if the input vertex is penetrated in the other object, we get all the vertices in the other object which have the shortest distance to this input vertex. In case the vertex is not penetrated inside the other object, Cv() returns an empty result.

## 2.3 Vertices of the network

Vertices of the network are exactly the vertices of the objects. Therefore, we define a set of network vertices corresponding to 3D vertices of the objects. The vertices can also include internal vertices of the object and not only surface vertices.

$$V = \{v_1, v_2, \ldots, v_N\}; \text{ N: total number of vertices}$$
$$v_i \longleftrightarrow a_i$$

## 2.4 Edges of the network

Edges are the connection between corresponding 3D vertices of the network vertices. That means if $a_i$ is connected to $a_j$ in the space, then $v_i$ is also connected to $v_j$ in the network. Therefore, there is an edge between vertex i and vertex j, if vertex i is connected to the vertex j. There are two kinds of edges in the network. First type connects the vertices of the same object to each other (we call them internal edges) and the other one connects the vertices of different object to each other (we call them external edge).

$$E = \{\varepsilon_1, \varepsilon_2, \ldots, \varepsilon_M\}; \text{ M: total number of edges}$$

### 2.4.1 Internal edges

Internal edge can be defined in different ways. One definition is that we consider vertex i is connected to vertex j if the Euclidian distance between them is less than a specific threshold. Another definition can be based on the object primitives (polygons or polyhedral). Therefore, we consider the vertex i connected to the vertex j if vertex i and vertex j are vertices of the same polygon (or polyhedral). For example, in Figure 1, two vertices of the surface are considered connected if they are vertices of the same triangle. Internal edges are used for transferring the internal forces between the primitives of an object.

### 2.4.2 External edges

For each vertex in object 1, there are edges connecting this vertex to all the vertices of the object 2. Thus, if object 1 and object 2 have $N_1$ and $N_2$ vertices, respectively, then the total number of external edges is $2 \times N_1 \times N_2$ (factor 2 is because of two directions). External edges are used for transferring the forces between two contacting objects.

## 2.5 The constant information stored for each vertex

For each vertex, we have some constant information which is not going to be changed. Therefore, we do not include them inside the state space of the vertices:

1- "The default 3D position of each vertex in the space, before any deformation".
For vertex v, we will have:

$$P^{(v)} = [p_x \ p_y \ p_z]^T$$

where $p_x$, $p_y$ and $p_z$ are components of v in the Cartesian coordinate system.

2- "The mobility status of the vertex"
It is possible that a vertex be fixed in its place, and never changes its position.

$$\text{Status}^{(v)} = \{0,1\}$$

1: if vertex v can change its position,
0: if vertex v is fixed in its position.

## 2.6   The state of the vertices

The state of each vertex has four parts:

$$\chi^{(v)} = \left\{\chi_1^{(v)}, \chi_2^{(v)}, \chi_3^{(v)}, \chi_4^{(v)}\right\}.$$

1- The current 3D Cartesian coordinate of each vertex is forming the first part of its state. Therefore, for vertex v, we will have:

$$\chi_1^{(v)} = [v_x \; v_y \; v_z]^T$$

2- The second part of the state is a set of 'corresponding vertices of vertex v, in the other object(s)'. Knowing about the corresponding vertices is necessary, as we need to balance the mutual forces imposed by object 1 on object 2 (and vice versa). The corresponding vertices may be changed during the simulation and needs to be updated in each step by using CollisionDetection() function.

$$\chi_2^{(v)} = \{ \text{ set of 'corresponding vertices of } v \text{ (in the other object(s))'} \} = Cv(v)$$

3- The third part of the state indicates the number of times that $\chi^{(v)}$ has been updated before.

$$\chi_3^{(v)} = \{0,1,2,\dots\}$$

4- The forth part of the state includes the first and second parts of the state of the neighboring vertices (the vertices which are directly connected to this vertex). This information of the neighboring vertices is needed for updating the signals of the edges.

$$\chi_4^{(v)} = \{ (\chi_1^{(vi1)}, \chi_2^{(vi1)}), \ (\chi_1^{(vi2)}, \chi_2^{(vi2)}), \dots \} ; \; v_{i1}, v_{i2}, \dots \text{ are connected to } v \text{ via one edge}$$

## 2.7   The signals of the edges

If a pair of vertices is connected, then they may impose force on each other. The signal of the edge connecting these two vertices contains the amount of the mutual force imposed on them.

$$s^{(e)} = \begin{bmatrix} F_{x,e} \\ F_{y,e} \\ F_{z,e} \end{bmatrix}$$

where $F_x$, $F_y$ and $F_z$ are the force elements in the x, y and z direction. If the direction of e is toward a vertex, it means the force is applying on that vertex.

## 2.8   External input

External input is sum of the forces applied on the objects from all the external sources, such as gravity.

$$s^{(v,in)} = \begin{bmatrix} F^{(v)}{}_{x,external} \\ F^{(v)}{}_{y,external} \\ F^{(v)}{}_{z,external} \end{bmatrix}$$

## 2.9   The state transition function

The stress-strain relation in a simple isotropic object can be written as:

$$\begin{bmatrix} \epsilon_x \\ \epsilon_y \\ \epsilon_z \end{bmatrix} = \left(\frac{1}{E}\right) \begin{bmatrix} 1 & -h & -h \\ -h & 1 & -h \\ -h & -h & 1 \end{bmatrix} \begin{bmatrix} \sigma_x \\ \sigma_y \\ \sigma_z \end{bmatrix}$$

where $h, E, \epsilon_i$ and $\sigma_i$ are Poisson's ratio, Young's modulus, the strain in i direction and the stress in i direction. The formulation can be modified in the below way to have a relation between force and displacement:

$$\begin{bmatrix} \Delta x \\ \Delta y \\ \Delta z \end{bmatrix} = \left(\frac{1}{k}\right) \begin{bmatrix} 1 & -h & -h \\ -h & 1 & -h \\ -h & -h & 1 \end{bmatrix} \begin{bmatrix} F_x \\ F_y \\ F_z \end{bmatrix} = K_{inv} \begin{bmatrix} F_x \\ F_y \\ F_z \end{bmatrix}$$

$$\begin{bmatrix} F_x \\ F_y \\ F_z \end{bmatrix} = \left(\frac{k}{1-h-2h^2}\right) \begin{bmatrix} 1-h & h & h \\ h & 1-h & h \\ h & h & 1-h \end{bmatrix} \begin{bmatrix} \Delta x \\ \Delta y \\ \Delta z \end{bmatrix} = K \begin{bmatrix} \Delta x \\ \Delta y \\ \Delta z \end{bmatrix}$$

$$k = EA_0/L_0$$

where $\Delta x$, $\Delta y$ and $\Delta z$ are amount of the displacement in x, y and z direction. $A_0$ is the cross-sectional area of the object primitives through which the force is applied, and $L_0$ is the length of the of the object primitive. Therefore, the state space of each vertex can be updated as below:

$$\chi_1^{(v)}(t') = \chi_1^{(v)}(t) + Status^{(v)} \times \alpha\, K_{inv} \left[ s^{(v,in)} + \sum_{\substack{for\ all\ the\ edges\ ei \\ directed\ toward\ v}} s^{(ei)} \right]$$

$$\alpha = (\gamma)^{\beta \chi_3^{(v)}+1}, \qquad 0 \le \beta \le 1\ ;\ 0 < \gamma < 1$$

$\beta$ and $\gamma$ must be decided in the way to help the network converge by an acceptable rate. After updating all the $\chi_1^{(v)}$, the collision detection program restarted again. The collision detection program finds the current penetrating vertices by using updated $\chi_1^{(v)}(t')$:

$$\text{Cv}(V) = \text{CollisionDetection}\left( \left\{ \chi_1^{(v1)}(t'), \chi_1^{(v2)}(t'), \dots, \chi_1^{(vN)}(t') \right\} \right)$$

$$\chi_2^{(v)}(t+1) = \text{Cv}(v)$$

Then, $\chi_1^{(v)}$ is re-updated by adding half of the penetrating depth to it:

$$\chi_1^{(v)}(t+1) = \chi_1^{(v)}(t') + \frac{1}{2} \times \text{Status}^{(v)} \times \left( -\chi_1^{(v)}(t') + \frac{1}{\#\{\chi_2^{(v)}(t+1)\}} \sum_{\substack{for\ all\ the \\ w \in \chi_2^{(v)}(t+1)}} \chi_1^{(w)}(t') \right)$$

If the objects have different mechanical characteristic, then instead of $\frac{1}{2}$, we should use $\frac{k1}{k1+k2}$ and $\frac{k2}{k1+k2}$ for the vertices of object 2 and the vertices of object 1, respectively. When the updating process is done, $\chi_3^{(v)}$ is also updated by adding a unit to it:

$$\chi_3^{(v)}(t+1) = \chi_3^{(v)}(t) + 1$$

## 2.10  Initial state

Both $\chi_1^{(v)}(0)$ and $\chi_2^{(v)}(0)$, are initialized by an external collision detection program. The collision detection program finds the penetrating vertices and the corresponding vertices of them in the other object(s):

$$\text{Cv}(V) = \text{CollisionDetection}\left( \{ \text{P}^{(v0)}, \text{P}^{(v1)}, \dots, \text{P}^{(vN)} \} \right)$$

$$\chi_2^{(v)}(0) = \text{Cv}(v)$$

$$\chi_1^{(v)}(0) = \text{P}^{(v)} + \frac{1}{2} \times \text{Status}^{(v)} \times \left( -\text{P}^{(v)} + \frac{1}{\#\{\chi_2^{(v)}(0)\}} \sum_{\substack{for\ all\ the \\ w \in \chi_2^{(v)}(0)}} \text{P}^{(w)} \right)$$

Again, if the objects have different mechanical characteristic, then instead of $\frac{1}{2}$, we should use $\frac{k1}{k1+k2}$ and $\frac{k2}{k1+k2}$ for the vertices of object 2 and the vertices of object 1, respectively. Since no updating process is done yet:

$$\chi_3^{(v)}(0) = 0.$$

## 2.11 Output function

For an internal edge 'e', $I(e) = (v_n, v_m)$, its signal is updated like below:

$$s^{(e)} = \left(\frac{1}{2}\right)^{\left(\text{Status}^{(vn)} \times \text{Status}^{(vm)}\right)} \times K\left(\left(\chi_1^{(v_m)} - P^{(v_m)}\right) - \left(\chi_1^{(v_n)} - P^{(v_n)}\right)\right)$$

The factor of $\left(\frac{1}{2}\right)$ is considered when both of the vertices are mobile (Status$^{(v1)}$ = Status$^{(v2)}$ = 1) and therefore the force is divided between them. In case both of the vertices are fixed $\left(\left(\chi_1^{(vm)} - P^{(vm)}\right) - \left(\chi_1^{(vn)} - P^{(vn)}\right)\right)$ will be zero.

For an external edge 'e', $I(e) = (v_n, v_m)$, its signal is updated like below:

$$s^{(e)} = \omega \times \sum_{\substack{\text{for all the internal edges } ej \\ \text{directed toward } v_m}} \frac{s^{(ej)}}{\#\left\{\chi_2^{(v_m)}\right\}}$$

$$\text{if } \left\{v_m \in \chi_2^{(v_n)}\right\} \text{ then } \omega = 1 \text{ ; else } \omega = 0$$

$$\#\{X\} \text{ is number of members of X}$$

As it can be seen, we use the state of both of the vertices of an edge for updating the edge signal. It is not against the fact that the signal should be updated based on the state of the vertex that the edge is originating from. The reason is that we have considered that the fourth part of the state contains the first and second parts of the state of the neighboring vertices (see section 2.6). Therefore, the first and second parts of the state of $v_n$ are stored inside the fourth part of the state of $v_m$ too.

## 2.12 Termination condition

The network needs a condition for terminating the updating process. The condition can be when the penetration depth of all the vertices is zero (ideally) or less than a pre-defined threshold ($\varepsilon_1$):

$$\left(-\chi_1^{(v)} + \frac{1}{\#\left\{\chi_2^{(v)}\right\}} \sum_{\substack{\text{for all the} \\ w \in \chi_2^{(v)}}} \chi_1^{(w)}\right) < \varepsilon_1 \text{ , For all } v \in V$$

and also, the amount of the total forces on all the vertices is zero (ideally) or less than a pre-defined threshold ($\varepsilon_2$):

$$\left( s^{(v,in)} + \sum_{\substack{for\ all\ the\ edges\ ei \\ directed\ toward\ v}} s^{(ei)} \right) < \varepsilon_2 \ , \quad \text{For all } v \in V$$

(We can also put a limit for the maximum number of updating (i.e. $\chi_3^{(v)}$)).

## 3   Implementation and results

We implemented the explained network in C++. For testing the network, we considered a sphere with radius of 45 units penetrating a cube with edge of 90 units. The surface of sphere is made up of 1026 points-2048 triangles, and the surface of cube is made up of 1538 points-3072 triangles (see Figure 4).

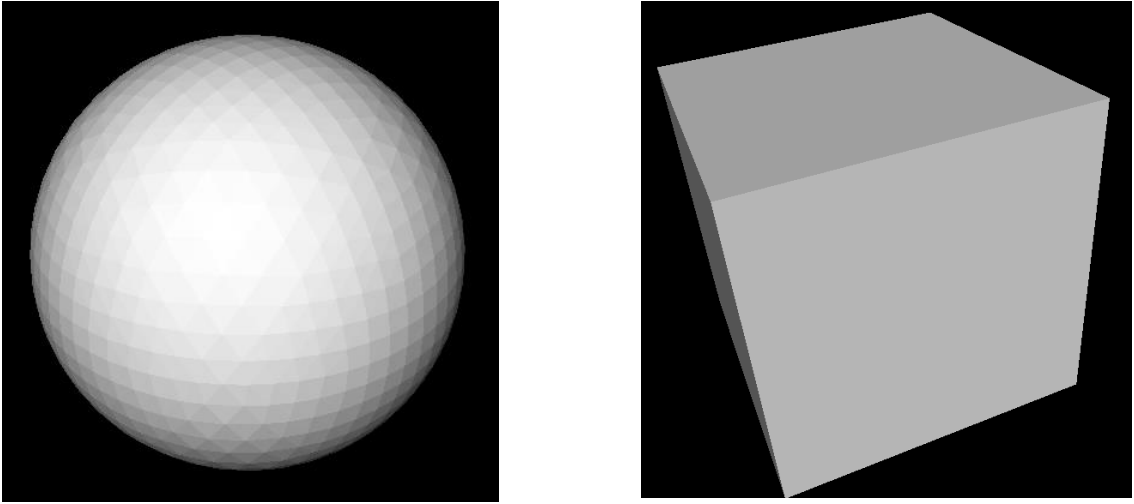

Figure 4: The sphere and cube used for testing the proposed dynamic network.

The sphere is penetrating about 17% of its diameter inside the cube. Half of the sphere and half of the cube are considered fixed and non-deformable (in Figure 5 these half parts are the upper and lower parts of the cube and the sphere, respectively). These fixed parts should be defined in order to avoid the objects get separated during the simulation. It is almost similar to assuming that the upper part of the cube and the lower part of the sphere are hold by two hands.

For having better tests, we also considered some points inside the objects. Without considering such internal points, the objects may behave like an empty box and an empty ball. An independent collision detection module was giving all the penetrating vertices, and their corresponding vertices, i.e. Cv(). In this module, the corresponding vertices of each penetrating vertex were in fact the closest vertices to it in the radial direction (from the center of sphere).

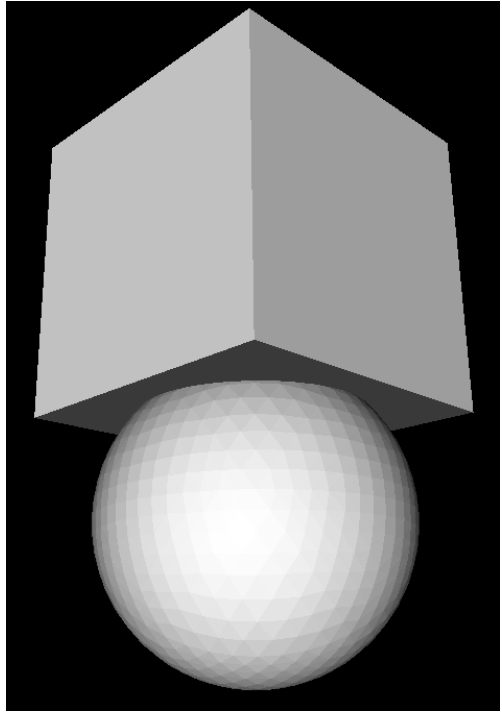

Figure 5: The sphere is penetrating the cube.

In section 2.9, we defined α based on γ and β, for making the network converge. In our experiment, choosing a large γ could cause an unstable situation with large unrealistic deformations. The reason is that having large γ may cause nonrealistic amount of displacement based on even small forces. Such a nonrealistic displacement can create much larger amount of force (to be applied on the vertices) and therefore a positive feedback can happen and makes the network fail. On the other hand, having a too small γ can make the process too slow. Finally, in our case we found γ = 0.1 to works almost better than the other values. Our experiments showed that having small β can make the network converge faster. Therefore, we chose a constant value of α = 0.1 for the network.

The amounts of k and h, depends on the mechanical characteristic of the objects and also geometrical characteristics of the object primitives (i.e. geometrical arrangement of vertices in the space). Usually the ratio of $k_1/k_2$ is more important than their individual values. It should be noticed that if $k_1 > k_2$, it does not necessarily mean that the object 1 is harder than the object 2. Such conclusion can be done, only if the primitives of the objects 1 and 2 are arranged exactly same as each other. We tested the network for different $k_1/k_2$ and h. The network terminated when the total delta-forces and total amount of penetration get smaller than the pre-defined values. The result related to one of the scenarios is shown in Figure 6.

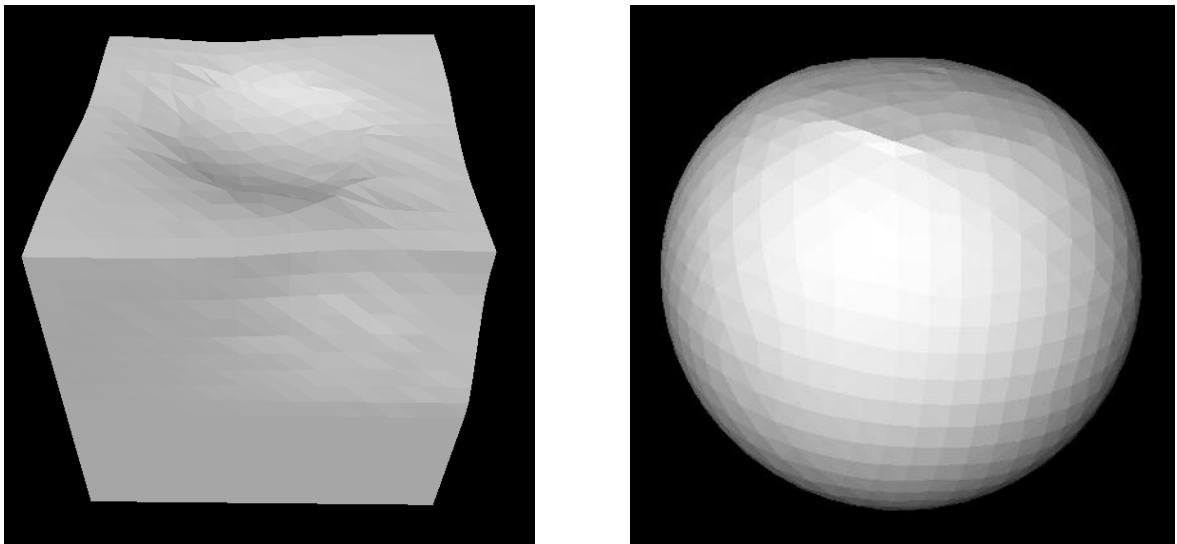

Figure 6: Deformation of cube and sphere, $k_{sphere}/k_{cube} = 3$ & h=0.

Since all of the nodes have interaction with each other (since all are part of one network), the non-penetrated vertices change their positions during the deformation too. In fact, the amount of deformation in a vertex is depending on the distance between this vertex and the penetrated vertices. This fact could be seen clearer when the sphere was considered as a rigid object comparing to the cube, (see Figure 7). As it can be seen in Figure 7, although the network starts by displacing the penetrating vertices, but due to the network connection, the other vertices (such as the edges of the cube) are also deformed.

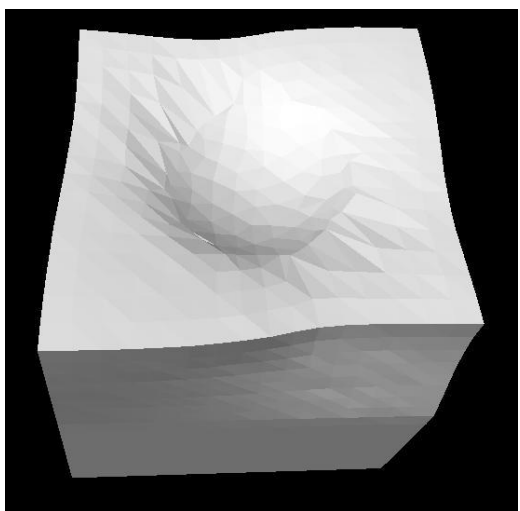

Figure 7: Deformable cube against the rigid sphere.

## 4   Conclusion

In this study, we tried to deform the penetrating objects, based on simple physics laws, by using the concepts of Dynamic Networks. We considered the geometrical vertices of the objects as the network nodes. The nodes of each object are connected to each other. Also, there are some connections between the nodes of different objects. The signal of each edge is in fact the force value transferred from one node to another. This force could change the state of the nodes (which contains its position) and consequently change the amount of forces applied on each vertex. The network works based on the negative feedback between the displacement of vertices and the total force imposed on them. This negative feedback works until all the forces get balanced.

The network was implemented and tested, and the results seemed satisfying (although more comparison scenarios may be needed for more accurate judgment). One problem in the network was its slow rate of converging. This problem was because of a positive feedback that could happen between force and displacement for large amount of α. Thus, we had to use a small amount of α, which consequently decrease the rate of converging, but can avoid such destabilizing positive feedback. As a future work, one can choose α as large as possible, and make it smaller only when a positive feedback is happening inside the network. Thus, α can be also a dynamic value changing based on the behavior of the network (e.g. vertices state and edges signal).